# Quantum Neural Network Inspired Hardware Adaptable Ansatz for Efficient Quantum Simulation of Chemical Systems

Xiongzhi Zeng,[1] Yi Fan,[1] Jie Liu,[2] Zhenyu Li,*,[1,2] Jinlong Yang[1,2]

[1]Hefei National Research Center for Physical Sciences at the Microscale, University of Science and Technology of China, Hefei, 230026, China

[2] Hefei National Laboratory, University of Science and Technology of China, Hefei 230088, China

Email: zyli@ustc.edu.cn

**Abstract:**

The variational quantum eigensolver is a promising way to solve the Schrödinger equation on a noisy intermediate-scale quantum (NISQ) computer, while its success relies on a well-designed wavefunction ansatz. Compared to physically motivated ansatzes, hardware heuristic ansatzes usually lead to a shallower circuit, but it may still be too deep for an NISQ device. Inspired by the quantum neural network, we propose a new hardware heuristic ansatz where the circuit depth can be significantly reduced by introducing ancilla qubits, which makes a practical simulation of a chemical reaction with more than 20 atoms feasible on a currently available quantum computer. More importantly, the expressibility of this new ansatz can be improved by increasing either the depth or the width of the circuit, which makes it adaptable to different hardware environments. These results open a new avenue to develop practical applications of quantum computation in the NISQ era.

**Keywords:**

hardware adaptable ansatz, low-depth circuit, expressibility, electronic structure theory

**Introduction**

With its advantage over classical computation being demonstrated recently,[1–3] quantum computation has attracted intense research interest. Solving electronic structure problems, where the main challenge is the exponentially increased Hilbert space with the number of electrons, is believed to be a killer application of quantum computation.[4–7] To exploit the quantum advantage, a well-designed quantum algorithm is required.[8] Quantum phase estimation is a powerful quantum algorithm for electronic structure calculations.[9,10] However, it requires a deep quantum circuit and thus a long coherence time, which is not realistic for near-term noisy intermediate-scale quantum (NISQ) devices.[11]

A more practical way to perform quantum computation on NISQ devices is using a hybrid quantum-classical algorithm. Variational quantum eigensolver (VQE) is such an algorithm to estimate the ground-state energy of a given Hamiltonian, based on a wavefunction ansatz or, more specifically, a parameterized quantum circuit.[12] The corresponding parameters are optimized on a classical computer with the energy and sometimes its gradient provided by a quantum computer.[13] The ansatz in a VQE algorithm can be chosen in many possible ways, and its quality determines the performance of the algorithm.

There are mainly two types of design strategies for wavefunction ansatzes in VQE.[5] The first is physically motivated. For example, in the unitary coupled cluster (UCC) ansatz, electron excitations from one to another orbital are considered. Although difficult on a classical computer, preparing a UCC state is convenient on a quantum computer. As a result, the UCC ansatz has been widely used in both theoretical and experimental studies.[12,14–20] Other physically motivated ansatzes can be constructed using the adaptive derivative-assembled pseudo-trotter (ADAPT) method,[21] the Hamiltonian variational method,[22] and the qubit coupled-cluster method.[23] In contrast to physically motivated ansatzes which systematically approximate the exact electronic wavefunction, hardware heuristic ansatzes focus on the quantum circuit itself, aiming to generate highly entangled states with low-depth circuits.[24–26] It has been demonstrated that hardware heuristic ansatzes, such as the hardware efficient ansatz (HEA),[24] can efficiently represent the solution space for electronic structure problems.

An HEA circuit contains multiple layers of simple circuit units. From the quantum resource point of view, the required number of layers determines the circuit depth. Another important quantum resource consideration is the number of qubits, which corresponds to the width of the circuit. In quantum computation, it is always desirable to have a circuit with both a low depth and a small width. However, due to the diversity of potential quantum computation platforms (superconducting, trapped-ion, photonic, etc),[2,3,27] the bottleneck of quantum resources in a specific case can be either the circuit depth or the width. For example, hundreds of qubits can be realized on a superconducting quantum computer, while the coherence time is limited to ~100 microseconds.[3,28] In contrast, the coherence time can reach dozens of seconds in a trapped-ion system, which is however hard to scale up.[29,30] Therefore, it is desirable to have a hardware adaptable ansatz with the flexibility in choosing its circuit with a depth or width priority. Notice that such a flexibility exists in artificial neural networks,[31] which can be used to prevent the network depth from increasing too quickly for complicated problems.[32]

In this study, we propose a general framework to design a hardware adaptable ansatz. Considering the width-depth flexibility in artificial neural networks, we start from the deep quantum neural network (QNN) model,[33] which is the quantum counterpart of the classical artificial neural network. In addition to the qubits representing the initial state, a large number of ancilla qubits are used in QNN. The number of required ancilla qubits increases quickly with the number of QNN layers. To make the number of qubits and the number of layers two independent model hyperparameters, we introduce the qubit reuse technique[34] in QNN. The obtained qubit-reuse QNN (qrQNN) circuit can be constructed with a fixed number of ancilla qubits independent of the number of QNN layers. The qrQNN circuit can be directly used as a wavefunction ansatz. However, qubit reuse requires measurements in the middle of the circuit, which is not convenient on some quantum computation platforms. By simplifying the qrQNN circuit, we propose a practical hardware adaptable ansatz (HAA) model. It turns out that HAA shows comparable expressive power to qrQNN. Compared to HEA, HAA requires much fewer circuit layers. At the same time, HAA has the flexibility to improve its expressibility by either increasing the number of layers (depth) or the number of ancilla qubits (width), which provides a powerful technique for electronic structure calculations in the NISQ era.

**Method**
**A. The hardware adaptable ansatz**

The Hamiltonian of a many-electron system in the formalism of second quantization is[4]

$$H = \sum_{ij} T_{ij} c_i^\dagger c_j + \frac{1}{2} \sum_{ijkl} V_{ijkl} c_i^\dagger c_j^\dagger c_k c_l \tag{1}$$

where $c_i^\dagger$ and $c_j$ are fermionic creation and annihilation operators associated with a single-electron orbital. $T_{ij}$ and $V_{ijkl}$ are one- and two-electron integrals. With Jordan-Wigner[35] or Bravyi-Kitaev[36] mapping, it can be transformed into a series of sums of Pauli matrices in the qubit space.

A parameterized wavefunction on a quantum computer can usually be written as

$$|\Psi\rangle = U(\Theta)|\Psi_0\rangle = \prod_{i=1}^{N} U_i(\theta_i)|\Psi_0\rangle \tag{2}$$

where $|\Psi_0\rangle$ is a reference state such as the Hartree-Fock state. $U_i(\theta_i)$ can either be associated with a specific electron excitation or correspond to a set of quantum gates depending on whether a physically motivated or a hardware heuristic ansatz is adopted. In both cases, the ground-state energy can be estimated by variationally minimizing the expected value of the Hamiltonian.

$$\min(E(\Theta)) = \min(\langle\Psi_0|U(\Theta)^\dagger H U(\Theta)|\Psi_0\rangle) \tag{3}$$

If $U_i(\theta_i) = \exp(-i\theta_i V_i)$, the gradients can be written as

$$\partial_k E = \frac{\partial E}{\partial \theta_k} = i\langle\Psi_0|U_-^\dagger [V_k, U_k^\dagger U_+^\dagger H U_+ U_k] U_-|\Psi_0\rangle \tag{4}$$

where $U_- = \prod_{i=0}^{k-1} U_i(\theta_i)$ and $U_+ = \prod_{i=k-1}^{N} U_i(\theta_i)$.

If ancilla qubits are used, a more general framework of quantum circuit-represented states can be constructed by introducing density matrices. Using QNN as an example, information propagates from the (*l*-1)th layer to the *l*'th layer via projecting out the (*l*-1)th layer states after a unitary operation on these two layers (Figure 1a).[33] Such a process can be mathematically described using density matrices

$$\rho^l = \text{tr}_{l-1}(U(\rho^{l-1} \otimes |0..0\rangle_l \langle 0..0|)U^\dagger) \tag{5}$$

where $\rho^l$ is the density matrix of the *l*'th layer and $\text{tr}_{l-1}$ means partial trace over the (*l*-1)th layer. A problem with such a protocol is that an additional set of qubits is required when an additional QNN layer is added. Therefore, the circuit width increases with its length. To solve this problem, the qubit reuse technique can be adopted, which leads to a qrQNN circuit with a fixed number of ancilla qubits (Figure 1b). Note that the qrQNN circuit is already similar to the HEA circuit (Figure 1c), which is also composed of multiple layers of circuit fragments with the same structure.

Qubit reuse is realized via middle-circuit measurement and reinitialization, which are not convenient on some quantum computer platforms. Since the much simpler HEA has been demonstrated to be capable of representing the wavefunction of molecules, we propose to simplify the qrQNN circuit by removing all middle-circuit measurements and reinitialization and tracing out ancilla qubits at the end of the circuit. Such a simplification leads to a new ansatz named HAA (Figure 1d). Mathematically, it can be written as

$$\rho^{out} = \text{tr}_{anc}\big(U(\rho^{in} \otimes \rho^{anc})U^\dagger\big) \qquad (6)$$

where $\rho^{anc} = |0..0\rangle\langle 0..0|$. The unitary $U$ is composed of $L$ layers of elementary unitary operations

$$U = \prod_{l=1}^{L} \prod_{i=1}^{N} U_i^l(\theta_i^l) \qquad (7)$$

We use HAA($n$, $L$) to denote an HAA circuit with $n$ ancilla qubits and $L$ circuit layers. HEA can be considered as a special case of HAA without ancilla qubits. HEA($L$) is used to name an HEA circuit with $L$ layers.

In an HAA circuit with specific $n$ and $L$, we still have the flexibility to choose the elementary unitary operations or, more specifically, the fundamental gates acting on one or two qubits. For example, a U3 and CNOT gate combination (U3CX) is used for adjacent qubits (adjacent coupling) in HEA, while a canonical gate combination (CAN)[37] is used for any pair of system and ancilla qubits (cross coupling) in qrQNN. The CAN gate combination contains more elementary CNOT gates, which is chosen as the default option for HAA in this study. In the HAA(0, L) case without ancilla qubit, we use the CAN gate combination with adjacent coupling.

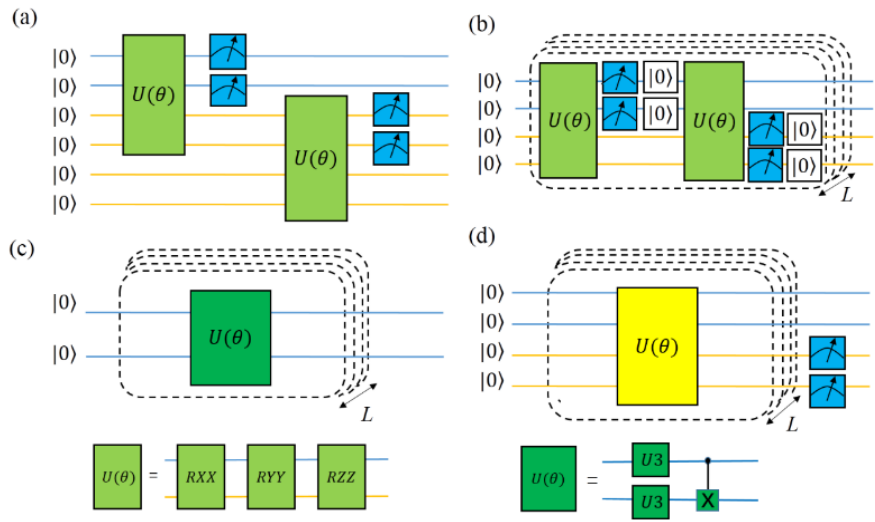

**Figure 1**. Schematics of quantum circuits for (a) QNN, (b) qrQNN, (c) HEA, and (d) HAA. System and ancilla qubits are marked with blue and yellow lines, respectively. A black dashed box indicates a circuit layer. The U3CX and CAN gate combinations are marked in green and light green,

respectively.

## B. Characterization of a parameterized quantum circuit

It is interesting to define some descriptors to evaluate different parameterized quantum circuits. One widely used descriptor is the expressibility,[38,39] which gives the capability of a circuit to generate quantum states. A parameterized quantum circuit can generate a state ensemble in the Hilbert space from a uniform sampling in the parameter space, while the distribution of states is usually not uniform. A larger deviation from the uniform state distribution leads to a low expressibility. Therefore, expressibility can be defined via the relative entropy or the Kullback-Leibler divergence ($D_{KL}$) of distributions of state fidelities referring to the uniform state distribution or the Haar state ensemble.[39,40] High expressibility ($D_{KL} \rightarrow 0$) means that the circuit can access more states in the Hilbert space and reach all the states with almost equal probabilities. Notice that a circuit with a high expressibility does not necessarily correspond to a good algorithm performance.[39]

VQE algorithms may encounter the vanishing gradient problem, also known as the barren plateaus problem,[41] which leads to a failure of the optimization of circuit parameters. Such a problem has been found in many parameterized quantum circuits, even the physical-inspired UCC quantum circuit.[42] Variance of the gradient can be used to characterize this phenomenon[41]

$$\text{Var}[\partial_k E] = \langle(\partial_k E)^2\rangle - \langle\partial_k E\rangle^2 = \langle(\partial_k E)^2\rangle \tag{8}$$

The average in the above definition is made for all possible $U$ operations and we choose the first variational parameter to calculate the variance of the gradient.[41]

## C. Numerical details

The VQE algorithm was simulated on classical computers using our homemade software Q$^2$Chemistry[43] with the source code deposited in gitlab.[44] The electronic structure parameters, including one and two electron integrals and the Hamiltonian operator, were calculated using the PySCF software.[45] Full configuration interaction (FCI) calculations were also performed using PySCF. Unless otherwise specified, the STO-3G basis set was used in all calculations.

## Results and Discussion

### 1. Expressibility and gradient variance

Before going to specific systems, we study the properties of HAA via some descriptors. To evaluate the expressibility, we calculate $D_{KL}$ for HAA($n$, $L$) circuits with four system qubits. $D_{KL}$ is small in all cases (Figure 2), which indicates that HAA has a high expressibility. As a comparison, the $D_{KL}$ of a four-qubit UCC circuit truncated at double excitation (UCCSD) for an H$_2$ molecule is 10.2 and that of its generalized UCCGSD version is still as high as 3.87. When $L$ is fixed, the experessibility of an HAA circuit can be improved by increasing $n$ (Figure 2a). When $n$ is fixed, expressibility can also be improved by increasing $L$ (Figure 2b). Therefore, there is a flexibility in HAA to improve its expressibility by either increasing the number of ancilla qubits or the number of circuit layers. Interestingly, although it is obtained by simplifying qrQNN, HAA has a higher expressibility compared to qrQNN for this four-qubit system. Compared to HEA, HAA with only one ancilla qubit can reach similar expressibility with a notably smaller number of circuit layers.

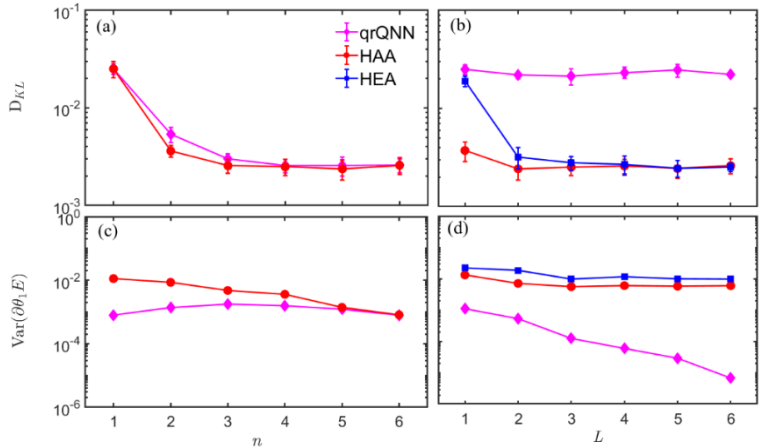

**Figure 2**. (a,b) Kullback-Leibler divergence and (c,d) gradient variance for a four-qubit system as functions of the number of ancilla qubits with the circuit layer number fixed to 1 and as functions of the number of circuit layers with the number of ancilla qubits in HAA and qrQNN fixed to 1.

Another measurement of an ansatz is the optimizability. A large gradient variance is helpful for parameter optimization in VQE algorithms. As shown in Figures 2c and 2d, a slight decrease in the gradient variance is observed in all circuits when the number of ancilla qubits or the number of circuit layers is increased. Compared to qrQNN, an HAA circuit with the same size is generally easier to optimize with a larger gradient variance. Compared to HEA, HAA is slightly more difficult to optimize since ancilla qubits are included. However, both HEA and HAA circuits have a significantly larger gradient variance compared to qrQNN with the same number of layers, and their decrease in gradient variance with the number of layers is much slower compared to qrQNN. Therefore, HAA is believed to have a high optimizability compared to other hardware-based models.

## 2. HAA for electronic structure calculations

We choose several molecular systems ($H_2$, LiH, $N_2$, and $H_2O$) to explore the power of the HAA quantum circuit. The hydrogen molecule with the minimal basis set has four spin orbitals, which can be represented with four qubits via Jordan-Wigner mapping. The number of qubits can be reduced to two by using the parity symmetry under the Bravyi-Kitaev mapping.[36] For such a simple system, the minimal HAA model with one ancilla qubit and one circuit layer already gives very accurate results. As shown in Figure 3a, the error of HAA(1,1) is already many orders of magnitude below the chemical accuracy (1 kcal/mol).

The LiH molecule has 12 spin orbitals and 10 qubits are used to represent its wavefunction via the Bravyi-Kitaev mapping. The HAA(1,8) ansatz can generate an accurate potential energy curve for this system. Here, the number of parameters increases to more than 240 with 8 entangling layers between the system qubits and the ancilla qubit. To ensure that the optimal parameters can be found, it is helpful to modify the loss function with physical constraints such as the conservation of particle number and total spin.[46] A more straightforward way is starting the optimization with dozens of initial parameters. In the LiH case, when necessary, we run 50 optimization jobs to search the desired state.

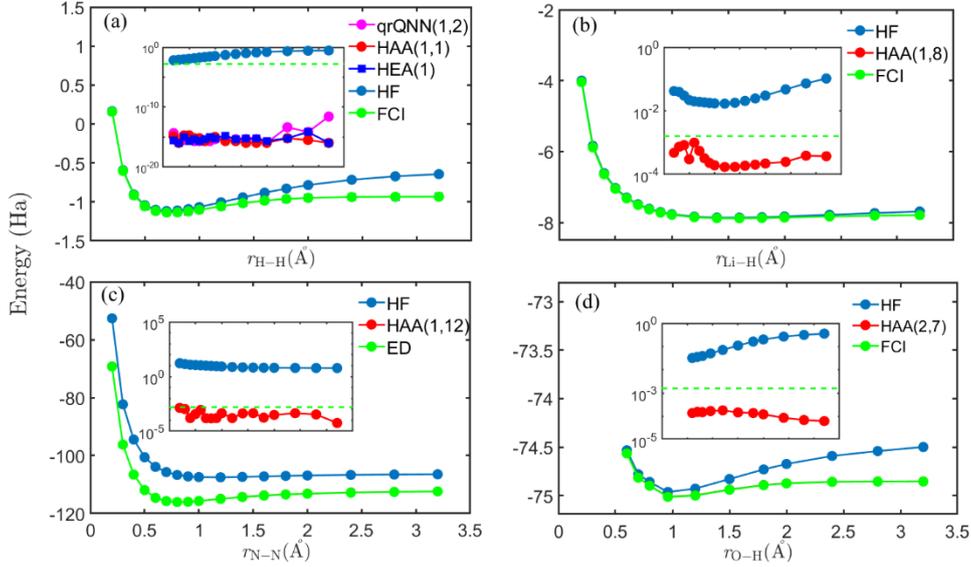

**Figure 3**. The dissociation potential energy of (a) $H_2$, (b) LiH, (c) $N_2$, and (d) $H_2O$. Inset: Absolute energy error compared to the exact-diagonalization FCI result. Green dashed lines mark the chemical accuracy. All energies are in Hartree.

The $N_2$ molecule is another example we used to test the HAA model. In this case, we freeze the lowest three core orbitals, which leads to 12 system qubits under the Bravyi-Kitaev mapping. As shown in Figure 3c, an HAA circuit with one ancilla qubit and 12 circuit layers is sufficient to give accurate ground-state energies compared to the exact diagonalization results. For the $H_2O$ molecule with 12 system qubits, an HAA(2,7) circuit leads to an accurate potential energy curve along the O-H distance. These results demonstrate that the HAA ansatz can be used in the VQE algorithm to solve the electronic structure problem of molecules.

3. **A comparison between HAA and HEA**

The HEA circuit has been successfully used to calculate the ground state energy of small molecules.[24] However, it is found that the circuit depth may increase rapidly even for relatively small systems.[5,24] We first check the $BeH_2$ system which has been simulated using a 28-layer HEA circuit in the literature.[24] With two frozen core orbitals, this system can be described by eight qubits. As shown in Figure 4a, in our test simulations using HEA circuits with different layer numbers, at least 25 layers are required to reach the chemical accuracy at all Be-H distances. When a single ancilla qubit is available, 8 layers are enough for an HAA circuit to reach the same goal. The depth of the HAA(1,8) circuit is much lower than that of the HEA(25) circuit. At the same time, the number of parameters to be optimized in the former (240) is also significantly smaller compared to that in the latter (600).

We use hydrogen chain systems to more systematically demonstrate the power of HAA to reduce the number of circuit layers by introducing ancilla qubits. For simplicity, we consider the cation for systems with odd numbers of hydrogen atoms. As shown in

Figure 4b, from the $H_2$ to $H_5^+$ chains, the number of HEA circuit layers required to obtain energies with the chemical accuracy increases from 1 to 21. Such a fast growth of the circuit depth of the HEA ansatz prohibits its application in relatively large molecular systems with NISQ hardware. When ancilla qubits are available (1, 2, 5, and 8 qubits for $H_2$ to $H_5^+$), a single layer of HAA circuit is enough to simulate these hydrogen chain systems within the chemical accuracy. Such a significant reduction in the circuit depth gives us great flexibility to choose proper quantum hardware.

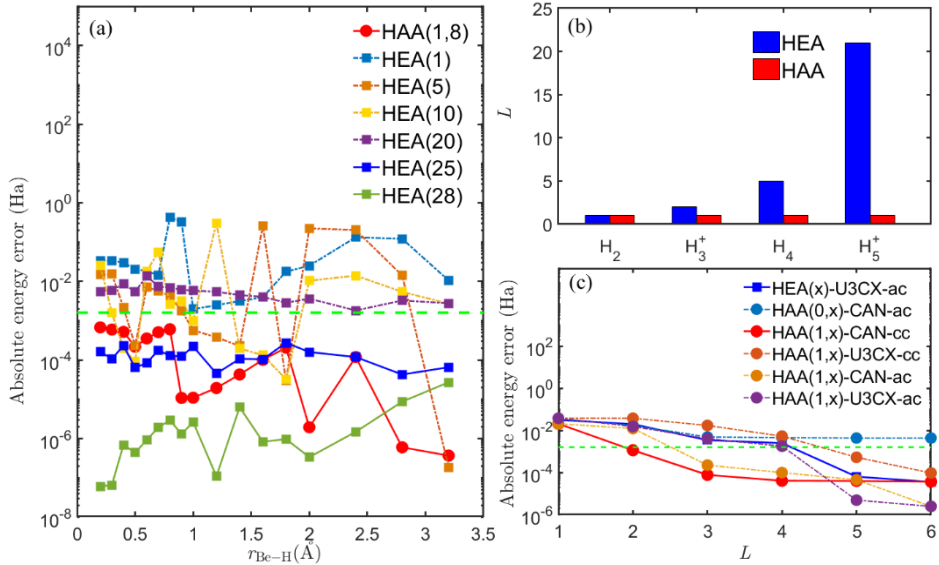

**Figure 4** (a) Absolute energy error of the $BeH_2$ molecule simulated with HEA and HAA circuits. (b) The minimum required number of layers for HEA/HAA to simulate H chain systems to reach the ground-state chemical accuracy. (c) Absolute energy error for the $H_4$ molecule at the equilibrium structure using different HAA and HEA circuits. Energies are in the unit of Hartree. The green dashed lines indicate the chemical accuracy.

Notice that the circuit units used in HEA and HAA are different. To check the effects of the circuit structure in each layer, we use the $H_4$ chain at its equilibrium structure as an example for a more detailed discussion. In each layer, we choose either CAN or U3CX gates with either adjacent coupling (ac) or cross coupling (cc). As shown in Figure 4c, the CAN-cc model we adopted for HAA indeed gives the overall best performance. When there is no ancilla qubit, the cross coupling mode does not apply. In this case, the HAA(0, $L$) circuit with the CAN-ac structure has an even worse performance compared to HEA($L$). Therefore, it is recommended to provide at least one ancilla qubit in the HAA circuit. An important conclusion from Figure 4c is that the ancilla qubit always improves the performance when the same circuit structure is used, which suggests that the idea of using ancilla qubits to improve the circuit performance is expected to be applicable to various circuit structures.

## 4.  Capability of HAA with a state-of-the-art quantum computer

Since HAA can significantly reduce the circuit depth by introducing ancilla qubits, it is interesting to check the capability of HAA on a state-of-the-art quantum computer. Although the number of available qubits in a quantum computer increases quickly, the number of qubits that can be used in a simulation is limited due to the presence of a coherence time limit and noise. For the same reason, there is also a tight restriction on how many quantum gates can be applied in a simulation. Another parameter that matters in experiment is the number of variational parameters. Too many variational parameters lead to low performance and difficulty in optimization. The maximum number of qubits used in reported quantum simulations of chemical systems is 16, where 160 Cz gates are applied with 160 parameters to be optimized.[47] Therefore, a simulation with 10-20 qubits using 150-250 Cz gates with 150-250 variational parameters represents the state of the art of current quantum simulation of chemical systems. Under such restrictions, most previous quantum simulations of chemical systems remain demonstrative, referring to an artificial theoretical model, for example, with a minimal basis set.[48–51] A practical quantum simulation should be able to generate data that can be tested with laboratory experiments.[52,53]

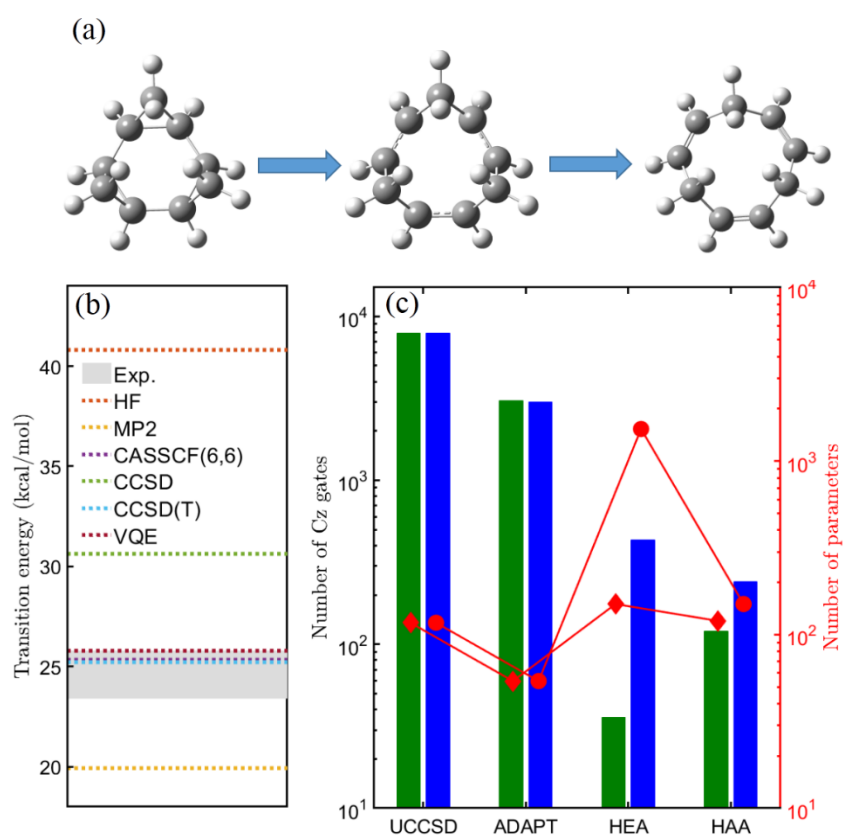

**Figure 5**. (a) Structures of the reactant, transition state, and product of the $C_9H_{12}$ cycloreversion reaction. Gray balls are carbon atoms and white balls are hydrogen atoms. (b) Activation energies predicted with different methods. The experimental values are within the gray region. (c) Required numbers of Cz gates and variational parameters in VQE simulations using UCCSD, ADAPT, HEA, and HAA ansatzes. The green and blue bars are for the reactant and the transition state, respectively.

As a demonstration, we try to use HAA to study the cycloreversion reaction of cis-triscyclopropacyclohexane ($C_9H_{12}$). Involving σ-bond breaking and π-bond formation, cycloreversion reactions are important in pharmaceutical syntheses.[54] The reactant, transition state, and product structures of the $C_9H_{12}$ cycloreversion reaction are shown in Figure 5a. This reaction has an experimental activation energy in the range of 23.4 to 25.8 kcal/mol.[55,56] Classical *ab initio* studies suggest that a CAS(6,6) model can accurately predict its activation energy.[57] Therefore, we build a 6-orbital model with a 6-31G(d) basis set and use the HAA ansatz to study this reaction. Converged results can be obtained by using 12-qubit HAA(2,1) and HAA(2,2) circuits for the reactant and the transition state, respectively. The obtained activation energy is 25.79 kcal/mol, which agrees well with the experimental results (Figure 5b). As shown in Figure 5c, the number of Cz gates used in the HAA(2,1) and HAA(2,2) circuits is 120 and 240, respectively, which is feasible for a state-of-the-art quantum computer. If we use physically motivated ansatz such as UCC, the circuit with thousands of Cz gates is too deep for current quantum devices. Even if the hardware efficient HEA ansatz is used, more than 400 Cz gates are required, which is currently challenging. Additionally, the number of variational parameters in HEA is much larger than that in HAA for this reaction. Therefore, although unfeasible with previous ansatzes, a practical quantum simulation of a chemical reaction with more than 20 atoms becomes possible now using the new HAA ansatz.

## 5. The hardware adaptability of HAA

The HAA ansatz has two important hyperparameters, the number of ancilla qubits and the number of circuit layers, which determine the width and depth of the corresponding circuit. In previous sections, we showed that fewer circuit layers are required in HAA with ancilla qubits compared to HEA without ancilla qubits. Now, we explore the relation between these two hyperparameters within the framework of HAA. We use the chain $H_4$ molecule as an example and perform VQE optimizations with different HAA($n$, $L$) circuits. As shown in Figure 6, a more accurate result converging towards the exact diagonalization result can be obtained with more ancilla qubits and circuit layers. There is a flexibility to reach the same accuracy with different circuits by either increasing the number of qubits or the number of layers. Such a flexibility gives the HAA model hardware adaptability. On quantum computation platforms where the number of qubits is difficult to increase, such as trapped-ion qubits,[58] an HAA circuit with a larger $L$ and smaller $n$ can be adopted. On platforms where increasing the coherence time is more challenging, such as a superconducting quantum computer,[28] an HAA circuit with a larger $n$ and smaller $L$ is desirable.

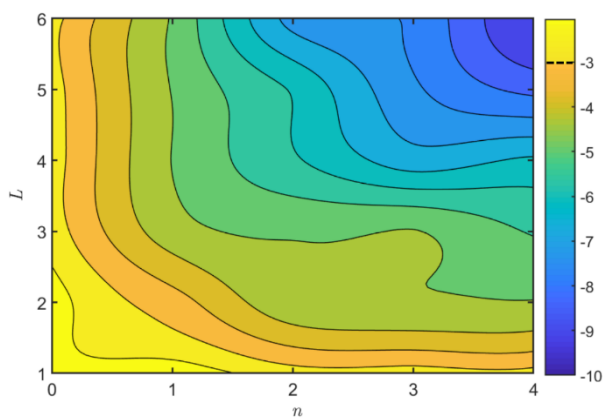

**Figure 6**. The highest accuracy reachable for VQE optimization of a chain $H_4$ molecule with an HAA($n$, $L$) circuit. The color bar is given in Hartree on a logarithmic scale where the chemical

accuracy is marked by a dashed line.

An interesting question here is whether different HAA circuits that give similar energy correspond to similar wavefunctions. With ancilla qubits introduced, the output state from an HAA circuit in principle can be a mixed state instead of a pure state. Of course, the ground state as a pure state is lower in energy compared to any mixed state. By checking if $\text{tr}((\rho^{\text{out}})^2) = 1$, we confirm that all those states we obtained within the chemical accuracy are pure states. We also compare these states with the accurate FCI state. For example, the two states predicted by HAA(1,6) and HAA(4,2) have similar energy accuracy ($1\times10^{-5}$ Hartree). As shown in Figure 7, their wavefunctions are already very similar to the FCI wavefunction as expected. The main difference comes from configuration number 28, where the HAA(1,6) and HAA(4,2) coefficients are notably smaller than the FCI coefficient. Such a difference disappears if a larger model, HAA(4,6), is adopted, which has an energy accuracy of $1\times10^{-10}$ Hartree.

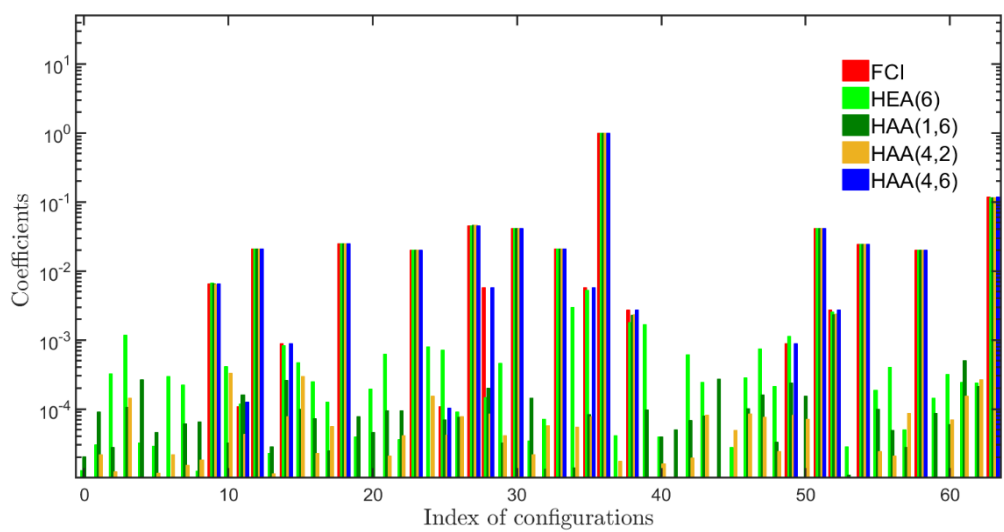

**Figure 7**. Configuration coefficients in the chain $H_4$ molecule wavefunction predicted by FCI and different HAA models.

**Conclusion**

In this study, inspired by the quantum neural network, we have designed a hardware adaptable ansatz to obtain the ground-state energy of chemical systems. The corresponding HAA quantum circuits demonstrate high expressibilities. By introducing ancilla qubits, the circuit depth and thus the number of variational parameters can be significantly reduced compared to that of the hardware efficient HEA ansatz. It is demonstrated that HAA can accurately predict the dissociation profile of small molecules. More importantly, it makes a practical simulation of a chemical reaction with more than 20 atoms on a currently available quantum computer possible. The expressibility of an HAA ansatz can be improved by increasing either the number of ancilla qubits (circuit width) or the number of entangled layers (circuit depth), according to the circuit optimization priority of a specific hardware platform. Therefore, HAA is expected to be a powerful wavefunction ansatz in VQE simulations of chemical systems in the NISQ era.

## Acknowledgement

This work is supported by the Innovation Program for Quantum Science and Technology (2021ZD0303306), the National Natural Science Foundation of China (21825302), the Fundamental Research Funds for the Central Universities (WK2060000018), and the USTC Supercomputing Center.